\documentstyle[12pt]{article}
\title{New results on the Stieltjes constants:  Asymptotic and
exact evaluation} 
\author{Mark W. Coffey\\
Department of Physics\\
Colorado School of Mines\\
Golden, CO  80401\\
(Received $\mbox{~~~~~~~~~~~~~~~~~~~~~~~~~~~~~~~2005}$)}
\date{April 9, 2005}
\begin{document}
\maketitle
\baselineskip=25 pt
\begin{abstract}

The Stieltjes constants $\gamma_k(a)$ are the expansion coefficients in
the Laurent series for the Hurwitz zeta function about $s=1$.
We present new asymptotic, summatory, and other exact expressions for 
these and related constants.  

\end{abstract}
 
\vspace{.25cm}
\baselineskip=15pt
\centerline{\bf Key words and phrases}
\medskip 

\noindent

Stieltjes constants, Riemann zeta function, Hurwitz zeta function, Laurent 
expansion, integrals of periodic Bernoulli polynomials, functional equation,
Kreminski conjecture

 
\baselineskip=25pt
\pagebreak
\medskip
\centerline{\bf Introduction}
\medskip

The Stieltjes (or generalized Euler) constants 
\cite{bala,briggs,dilcher,ivic,kluyver,kreminski,mitrovic,stieltjes,wilton,zhang}
$\gamma_k(a)$ appear as 
expansion coefficients in the Laurent series about $s=1$ of the Hurwitz 
zeta function $\zeta(s,a)$, a generalization of the Riemann zeta 
function $\zeta(s)$ \cite{berndt,edwards,ivic,karatsuba,riemann,titch}.
We present both new asymptotic and 
exact expressions for these and other fundamental mathematical constants.  

In the following, we let $s(n,m)$ denote the Stirling numbers of the first
kind and $C_k(a) \equiv \gamma_k(a) -(\ln^k a)/a$.  By convention, 
$\gamma_k$ represents $\gamma_k(1)$.  With $B_n(x)$ the
Bernoulli polynomials, their periodic extension is denoted 
$P_n(x) \equiv B_n(x-[x])$.  As is customary, $B_j$ denotes the Bernoulli
numbers $B_j(0)=(-1)^jB_j(1)$.

{\bf Proposition 1}.  As $n \to \infty$, we have $C_n(a+1/2) \to -C_n(a)$.

{\bf Proposition 2}.  For integers $m>0$ and $n \geq 1$ we have 
$$\gamma_n=(-1)^{n-1}n!m^{1-n}\sum_{k=0}^{n+1}{{s(n+1,n+1-k)} \over {k!}}
\int_1^\infty P_n(mx) {{\ln^k x} \over x^{n+1}} dx
-\sum_{r=1}^{m-1} C_n(r/m).$$

{\bf Proposition 3}.  For integers $q \geq 2$ we have
$$\sum_{r=1}^{q-1} \gamma_k(r/q)=-\gamma_k + q(-1)^k {{\ln^{k+1} q} \over
{(k+1)}} + q\sum_{j=0}^k {k \choose j} (-1)^j (\ln^j q) \gamma_{k-j}.$$

{\bf Proposition 4}.  For $\sigma = $ Re $s<1$ we have
$$\sum_{k=0}^\infty {{(-1)^k} \over k!} (s-1)^k \int_0^1 \gamma_k(a) da
={1 \over {1-s}}, ~~~~~~ \sigma <1.$$

{\bf Proposition 5}.  For large but fixed $k$, $0 < a \leq 1$, we have
$$\gamma_k(a) \sim m \sin[2\pi(a+\phi)],$$
for real $m$ and $0 \leq \phi <1$.

{\bf Proposition 6}.  For integers $n \geq 1$ and $k \geq 1$ we have  
$$\sum_{j=0}^{n-1} C_k\left(a \pm {j \over n}\right )=(-1)^{k-1}k!n^{1-k}
\sum_{\ell=0}^{k+1} {{s(k+1,k+1-\ell)} \over {\ell!}}\int_1^\infty P_k[n(x-a)] 
{{\ln^\ell x} \over x^{k+1}} dx.$$

{\bf Proposition 7}.  For integers $j \geq 1$ we have the set of coupled
differential equations
$${{(-1)^j} \over {j!}}{{d\gamma_j(a)} \over {da}}=-\sum_{k=j-1}^\infty
{{(-1)^k} \over {k!}}{{k+1} \choose j}\gamma_k(a), ~~~~~~j \geq 1,$$
and
$$-\psi'(a)=-1-\sum_{k=0}^\infty {{(-1)^k} \over {k!}}\gamma_k(a),$$
where $\psi$ is the digamma function.

{\bf Proposition 8}.  We have (a)
$$\gamma_0=\gamma={1 \over 2}\ln 2-{1 \over {\ln 2}}
\sum_{n=0}^\infty {1 \over 2^{n+1}}\sum_{k=1}^n (-1)^k{n \choose k} {{\ln(k+1)} 
\over {(k+1)}},$$ 
for the Euler constant, (b)
$$-\gamma_1= {{\ln^2 2} \over {12}}-{1 \over 2}
\sum_{n=0}^\infty {1 \over 2^{n+1}}\sum_{k=1}^n (-1)^k{n \choose k} {{\ln(k+1)} 
\over {(k+1)}}+{1 \over {2\ln 2}}\sum_{n=0}^\infty {1 \over 2^{n+1}}
\sum_{k=1}^n (-1)^k{n \choose k} {{\ln^2(k+1)} \over {(k+1)}},$$
(c)
$$\eta_1={{\ln^2 2} \over {12}}+{1 \over 2}{1 \over {\ln^2 2}}
\left[\sum_{n=0}^\infty {1 \over 2^{n+1}}\sum_{k=1}^n (-1)^k{n \choose k} 
{{\ln(k+1)} \over {(k+1)}}\right]^2-{1 \over {\ln 2}}\sum_{n=0}^\infty 
{1 \over 2^{n+1}}\sum_{k=1}^n (-1)^k{n \choose k} {{\ln^2(k+1)} \over 
{(k+1)}},$$
(d)
$$\ln \pi = \ln 2 - 2\sum_{n=0}^\infty {1 \over 2^{n+1}}
\sum_{k=1}^n (-1)^k {n \choose k} \ln(k+1),$$
(e)
$$(-1)^{n-1}{{2^{2n-1}\pi^{2n}} \over {(2n)!}}B_{2n}=
{1 \over {1-2^{1-2n}}}\sum_{\ell=0}^\infty {1 \over 2^{\ell+1}}
\sum_{k=0}^\ell (-1)^k {\ell \choose k} {1 \over {(k+1)^{2n}}}$$
$$={1 \over {(2n-1)}}\sum_{\ell=0}^\infty {1 \over {\ell+1}}
\sum_{k=0}^\ell (-1)^k {\ell \choose k} {1 \over {(k+1)^{2n-1}}},
~~~~~~ n \neq 0,$$
and (f) the summation identity
$${1 \over {(2^{2n+1}-1)}}\sum_{n=0}^\infty {1 \over 2^{n+1}}
\sum_{k=1}^n (-1)^k {n \choose k} \ln(k+1) (k+1)^{2n}$$
$$=(-1)^n {{(2n)!} \over {2(2\pi)^{2n}}}{1 \over {(1-2^{-2n})}}
\sum_{n=0}^\infty {1 \over 2^{n+1}}
\sum_{k=0}^n (-1)^k {n \choose k} {1 \over {(k+1)^{2n+1}}}.$$
In part (c), the constants $\eta_j$ are 
defined by the expansion \cite{coffey,coffey03,coffeygam1}
$\ln \zeta(s)=-\ln(s-1)-\sum_{p=1}^\infty {\eta_{p-1} \over p}(s-1)^p$, 
$s \neq 1$.

These theorems are representative, but not comprehensive, of our results.
We do not reproduce relevant background material on properties and
functional equations of the special functions involved, but simply mention 
the useful sources \cite{nbs,grad,hansen,sri,zhang}.

\centerline{\bf Proofs of the Propositions}
\medskip

Here we summarize or otherwise simply indicate the method of proof of the
Propositions above.

Proposition 1.  A key observation is that
$$P_{2n}(x-a)=(-1)^{n-1}{{2(2n)!} \over {(2\pi)^{2n}}}\left[\cos 2\pi (x-a)
+ O(2^{-2n}) \right ], \eqno(1)$$
and similarly for $P_{2n+1}(x-a)$, based upon the Fourier expansions of
these polynomials \cite{nbs}.  So for large $n$, $P_{2n}(x-a) = -P_{2n}
(x-a-1/2)+O(2^{-2n})$ and we may appeal to the representation
\cite{zhang}
$$C_n(a) = (-1)^{n-1} n!\sum_{k=0}^{n+1} {{s(n+1,n+1-k)} \over {k!}}
\int_1^\infty P_n(x-a) {{\ln^k x} \over x^{n+1}} dx, ~~n \geq 1. \eqno(2)$$
Due to the boundedness of $P_n$, $|P_n(x)| \leq [3+(-1)^n]/(2\pi)^n$ for
$n \geq 1$ \cite{berndt}, there is uniform convergence of the integral in
this equation.  Therefore, we may use expressions like (1) in Eq. (2), 
interchange the limit and integration, and Proposition 1 follows.
We have thereby proved a recent Conjecture II put forth by Kreminski
\cite{kreminski}.  

Proposition 2.  The multiplication formula satisfied by the Bernoulli
polynomials \cite{nbs} carries over to $P_n$.  Applying this to the
representation (2) yields the Proposition.

Proposition 3.  One procedure is to take derivatives
of the special case of the Hurwitz zeta function 
$\sum_{r=1}^{q-1} \zeta(s,r/q)=(q^s-1)\zeta(s)$ \cite{hansen} 
and to evaluate as $s \to 1^+$.  Another way is to use
the defining Laurent expansions for the Stieltjes constants 
\cite{kreminski,stieltjes,zhang} in this special case, writing
$$\sum_{r=1}^{q-1}\left [{1 \over {s-1}}+\sum_{k=0}^\infty {{(-1)^k \gamma_k
(r/q)} \over k!} (s-1)^k\right] =\left[qe^{(s-1)\ln q} - 1\right]\zeta (s)$$
$$=\left[q\sum_{j=0}^\infty {{\ln^j q} \over {j!}} (s-1)^j-1\right ]
\left[{1 \over {s-1}}+\sum_{k=0}^\infty {{(-1)^k \gamma_k} \over
k!} (s-1)^k \right].  \eqno(3)$$ 
We expand the right side of this equation and cancel the term $(q-1)/(s-1)$
on both sides.  We then reorder the double sum on the right side and equate
coefficients of like powers of $s-1$ on each side of the resulting equation.
The Proposition follows once again.  Proposition 3 takes the same form as
Corollary 11 of Dilcher \cite{dilcher}, who introduced a generalized
digamma function.  We explain this more in the Discussion section.

Proposition 4.  The key here is to note that 
$$\int_0^1 \zeta(s,a) da = 0, ~~~~~~~~ \sigma < 1,  \eqno(4)$$
and then to integrate the Laurent expansion of the Hurwitz zeta function.
Remark.  Many other related integral-based results for the Stieltjes
constants are possible.

Proposition 5.  For proof, we give the details for $C_{2n-1}(a)$, 
with the case for $C_{2n}(a)$ being very similar.  From Eq. (2) we have
$$C_{2n-1}(a) = (2n-1)!\sum_{k=0}^{2n} {{s(2n,2n-k)} \over {k!}}
\int_1^\infty P_{2n-1}(x-a) {{\ln^k x} \over x^{2n}} dx, ~~n \geq 1. 
\eqno(5)$$
We have \cite{nbs}
$$P_{2n-1}(x-a)=(-1)^n{{2(2n-1)!} \over {(2\pi)^{2n-1}}}\left[\sin 2\pi (x-a)
+ O(2^{1-2n}) \right ], \eqno(6)$$
so that we obtain
$$C_{2n-1}(a)=\kappa_n \sum_{k=0}^{2n} {{s(2n,2n-k)} \over {k!}}\int_1^\infty
\left[\cos 2\pi a\sin 2\pi x-\sin 2\pi a\cos 2\pi x+O(2^{1-2n})\right]
{{\ln^k x} \over x^{2n}} dx,  \eqno(7)$$
where $\kappa_n \equiv (-1)^n2[(2n-1)!]^2/(2\pi)^{2n-1}$.
As a result of the integration in this equation, the leading term in
$C_{2n-1}(a)$ may be written as
$$C_{2n-1}(a) \sim r_1(n)\cos 2\pi a -r_2(n)\sin 2\pi a, \eqno(8)$$
where
$$r_1(n)=\kappa_n \sum_{k=0}^{2n} {{s(2n,2n-k)} \over {k!}}\int_1^\infty
\sin 2\pi x {{\ln^k x} \over x^{2n}} dx,  \eqno(9a)$$
and
$$r_2(n)=\kappa_n \sum_{k=0}^{2n} {{s(2n,2n-k)} \over {k!}}\int_1^\infty
\cos 2\pi x {{\ln^k x} \over x^{2n}} dx.  \eqno(9b)$$
The form (9) can always be written as given in the Proposition with
$\tan 2\pi \phi=-r_1/r_2$ and $m^2 = r_1^2 + r_2^2$.
Remark.  The integrals appearing in Eqs. (9) can be written in several
different ways via integration by parts, but we do not pursue this here.

Proposition 6 follows from Eq. (2), the multiplication formula satisfied
by $P_n$ \cite{nbs}, and the interchange of two finite sums.

Proposition 7.  
According to the property $\partial \zeta(s,a)/\partial a=-s\zeta(s+1,a)$
and the Laurent expansion of $\zeta(s,a)$ we have
$$\sum_{k=0}^\infty {{(-1)^k} \over k!} {{\gamma_k(a)} \over {da}}(s-1)^k
=-1-\sum_{k=0}^\infty {{(-1)^k} \over k!} \gamma_k(a) s^{k+1},
\eqno(10)$$
wherein $d\gamma_0(a)/da=-\psi'(a)$ and $\psi'$ is the trigamma function.
If we perform a binomial expansion on the right side of this equation, 
reorder the sums there, and then equate coefficients of like powers of $s-1$
on both sides, we arrive at the stated set of equations.

Proposition 8.
Valid in the whole complex plane is the form of the Riemann zeta function
$$\zeta(s)={1 \over {1-2^{1-s}}}\sum_{n=0}^\infty {1 \over 2^{n+1}}
\sum_{k=0}^n (-1)^k {n \choose k} {1 \over {(k+1)^s}}, ~~~~s \neq 1.
\eqno(11)$$
This expression, due to Hasse \cite{hasse}, can be derived by applying
Euler's series transformation to the alternating zeta function \cite{sondow}.

For purposes of expanding Eq. (11) about $s=1$ we have 
$$1-2^{1-s}=-\sum_{j=1}^\infty {{(-\ln 2)^j} \over {j!}} (s-1)^j,
\eqno(12a)$$
$$(k+1)^{-s}=\sum_{\ell=0}^\infty{{(-1)^\ell} \over {\ell!}} \ln^\ell(k+1)
\sum_{q=0}^\ell {\ell \choose q} (s-1)^q, \eqno(12b)$$
and 
$$[1-2^{1-s}]^{-1}={1 \over {\ln 2(s-1)}}+{1 \over 2}+{{\ln 2} \over {12}}
(s-1)- {{\ln^3 2} \over {720}}(s-1)^3+O[(s-1)^5].  \eqno(12c)$$

By using the series $\sum_{k=0}^n {n \choose k} {{(-1)^k} \over {(k+1)}}
={1 \over {n+1}}$ it is easy to see that the $q=0$ term of Eq. (12b)
contributes a $\ln 2$ term in Eq. (11):
$$\sum_{n=0}^\infty {1 \over 2^{n+1}}\sum_{k=0}^n {n \choose k} {{(-1)^k} 
\over {(k+1)}} =\ln 2.  \eqno(13)$$ 
Therefore, comparing with the Laurent expansion of $\zeta(s)$ we obtain 
parts (a) and (b).

Similarly, if we write
$$\ln \zeta(s)=-\ln(1-2^{1-s})+\ln\left[\sum_{n=0}^\infty {1 \over 2^{n+1}}
\sum_{k=0}^n (-1)^k {n \choose k} {1 \over {(k+1)^s}}\right ]$$
$$=-\ln(s-1)+{1 \over 2}\ln 2(s-1)+{1 \over {24}}(s-1)^2+O[(s-1)^4]$$
$$+\ln\left[\sum_{n=0}^\infty {1 \over 2^{n+1}}
\sum_{k=0}^n (-1)^k {n \choose k} \sum_{\ell=0}^\infty{{(-1)^\ell} \over 
{\ell!}} \ln^\ell(k+1) \sum_{q=1}^\ell {\ell \choose q} (s-1)^q\right ],
\eqno(14)$$
we find $-\eta_0=\gamma$ and the expression given in part (c).

From Eq. (11) we have
$$\zeta'(s)=-{{\ln 2} \over {(2^{s-1}-1)}}\zeta(s)
-{1 \over {1-2^{1-s}}}\sum_{n=0}^\infty {1 \over 2^{n+1}}
\sum_{k=1}^n (-1)^k {n \choose k} {{\ln(k+1)} \over {(k+1)^s}}, ~~~~s \neq 1.
\eqno(15)$$
Part (d) follows from the well known value
$\zeta'(0)/\zeta(0)=\ln 2\pi$.
Similarly, based upon the well known values of the zeta function at
positive even integers we obtain part (e) for the Bernoulli numbers.

From the functional equation of the zeta function comes the evaluation
for positive integers $n$
$$\zeta'(-2n)=(-1)^n {{(2n)!\zeta(2n+1)} \over {2(2\pi)^{2n}}}.  \eqno(16)$$
From the representation (11) and another globally but more slowly
convergent series for $\zeta(s)$ due to Hasse \cite{hasse} we then have the
identity of part (f).

\medskip
\centerline{\bf Discussion}
\medskip


We may note the special values of the Stirling numbers of the first kind
\cite{nbs} $s(n,0)=\delta_{n0}$ and $s(n,n)=1$.  Therefore, the general
integral representation of Eq. (2) can be simplified to
$$C_n(a) = (-1)^{n-1} n!\left[\int_1^\infty {{P_n(x-a)} \over x^{n+1}}dx
+ \sum_{k=1}^n {{s(n+1,n+1-k)} \over {k!}} \int_1^\infty P_n(x-a) 
{{\ln^k x} \over x^{n+1}} dx \right], ~~n \geq 1. \eqno(17)$$
When $a=1$, the first term on the right side of this equation may be 
evaluated by using integration by parts on the result of the Appendix.

Despite some of the complexity of the representation (11), we have 
several motivations for examining it.  (i)   This form gives rise to
series that play a significant role in renormalization and quantum
field theory \cite{bloch,el}.  
(ii) Equation (11) has connections to random variables in an analytic
number theory setting \cite{biane}.  There is a possibility in particular
to link this structure with that of Brownian processes.
This brings us to a third point of interest.  (iii)  We would like to 
know not only the structure of the Stieltjes constants but also of
the constants $\{\eta_j\}$ that enter the Laurent expansion of the
logarithmic derivative of $\zeta(s)$.  Such knowledge would be very
influential in deciding the nature of a sum denoted $S_2(n)$ 
\cite{coffey,coffey03,coffeygam1} and thereby
the resulting Li/Keiper constants $\lambda_k$ \cite{li,keiper}.

As a byproduct of this work we obtain interesting infinite series for
fundamental constants such as the Euler constant and $\ln \pi$.  The
rapidity of convergence may make some of these suitable for 
applications.  Indeed, even a naive numerical implementation of part (a)
or (b) of Proposition 8 appears to provide these constants to 16 decimal
places after summing over $n$ to $51$ or $52$ terms.

The generalized digamma function used by Dilcher \cite{dilcher} is given by
$$\psi_k(a) \equiv -\gamma_k-{1 \over a}\ln^k a-\sum_{\nu=1}^\infty
\left[{{\ln^k(\nu+a)} \over {\nu+a}}-{{\ln^k a} \over \nu}\right], ~~~~
k \geq 0, \eqno(18)$$
and we may relate it to other functions defined in the literature.
We have $R_m(a)=(-1)^{m+1}(\partial^m/\partial s^m)\zeta(0,a)$ \cite{kanemitsu}
and find that $\psi_k(a)=R'_{k+1}(a)/(k+1)$.  We have
$R_j(a)=(-1)^j j!-\sum_{\ell=0}^\infty \gamma_{j+\ell}(a)/\ell!$ and by using
our Proposition 7 we determine that $\psi_j(a) = -\gamma_j(a)$.  This relation
explains how our Proposition 3 takes exactly the same form as Dilcher's 
Corollary 11 and permits the re-expression of several of his other results as
well.

In his computational work based upon Newton-Cotes integration for the
high accuracy approximation of the Stieltjes constants, Kreminski observed
that for large values of $k$, $\gamma_k(1/2) \approx -\gamma_k$ and that
more generally $C_k(a+1/2) \approx -C_k(a)$.  That is, 
in a sense, the Stieltjes constants for large index are anti-periodic
with period $1/2$.  Our Proposition 1 makes this precise.

Much earlier, Hansen and
Patrick \cite{hansen} showed that a fundamental interval in $a$ for the 
Hurwitz zeta function need only be of length $1/2$.  Perhaps solely on
that basis one would then suspect that some sort of relationship(s) should
exist between $\gamma_k(a)$ and $\gamma_k(a+1/2)$.

Much of our development relies on the underlying theory of periodized Bernoulli
polynomials and corresponding integral representations of the 
Stieltjes constants \cite{zhang}.  Equations such as (17) and still others
that we have obtained help to expose more of the analytic structure of 
the Stieltjes and $\eta_j$ constants. 

\pagebreak
\centerline{\bf Appendix:  Integrals over periodic Bernoulli polynomials}

Herein we evaluate integrals over the periodic Bernoulli polynomial
$P_1$ in terms of polygamma functions $\psi^{(j)}$.  We demonstrate
{\newline \bf Proposition} for positive integers $n$ and $m$
$${{(-1)^n} \over {(n+1)!}} \psi^{(n)}(s)+{1 \over {(n+1)}}\sum_{k=0}^m
{1 \over {(x+k)^{n+1}}}=\int_m^\infty {{P_1(x)} \over {(x+s+1)^{n+2}}}dx
+{1 \over {(n+1)}}{1 \over {(s+m)^{n+1}}}$$
$$-{1 \over {n(n+1)}}{1 \over 
{(s+m+1)^n}}-{1 \over 2}{1 \over {(n+1)}}{1 \over {(s+m+1)^{n+1}}}.  
\eqno(A.1)$$

Our starting point is \cite{edwards}
$$\ln \Gamma(s+1)=(s+1/2)\ln s - s+{1 \over 2}\ln 2\pi-\int_0^\infty
{{P_1(x)} \over {(x+s)}} dx, \eqno(A.2)$$
so that
$$\psi(s+1)=\ln s+ {1 \over {2s}}+\int_0^\infty {{P_1(x)} \over {(x+s)^2}}dx,
\eqno(A.3)$$
where $\psi(s)=\psi(s+1)-1/s$ is the digamma function.  We now take $n$ 
derivatives of relation (A.3), obtaining
$${{(-1)^n} \over {(n+1)!}}\psi^{(n)}(s+1)=-{1 \over {n(n+1)}}{1 \over s^n}
+{1 \over 2}{1 \over {(n+1)}}{1 \over s^{n+1}}+\int_0^\infty {{P_1(x)} \over
{(x+s)^{n+2}}} dx.  \eqno(A.4)$$
Finding that
$$\int_0^\infty {{P_1(x)} \over {(x+s)^{n+2}}} dx =
\int_1^\infty {{P_1(x)} \over {(x+s)^{n+2}}}dx+{1 \over {n(n+1)}}\left[{1 
\over s^n}-{1 \over {(s+1)^n}}\right ]+{1 \over {2(n+1)}}\left[{1 \over s^{n+1}}
-{1 \over {(s+1)^{n+1}}}\right ], \eqno(A.5)$$
we obtain
$${{(-1)^n} \over {(n+1)!}}\psi^{(n)}(s+1)={1 \over {(n+1)}}{1 \over s^{n+1}}
-{1 \over {n(n+1)}}{1 \over {(s+1)^n}}-{1 \over 2}{1 \over {(n+1)}}{1 \over 
{(s+1)^{n+1}}}+\int_1^\infty {{P_1(x)} \over {(x+s)^{n+2}}}dx.  \eqno(A.6)$$

By making a simple change of variable in the integral of Eq. (A.6) and using
the periodicity of $P_1$, we have
$${{(-1)^n} \over {(n+1)!}}\psi^{(n)}(s+m+1)={1 \over {(n+1)}}{1 \over 
{(s+m)^{n+1}}}-{1 \over {n(n+1)}}{1 \over {(s+m+1)^n}}$$
$$-{1 \over 2}{1 \over {(n+1)}}{1 \over {(s+m+1)^{n+1}}}+\int_m^\infty 
{{P_1(x)} \over {(x+s+1)^{n+2}}}dx.  \eqno(A.7)$$
Applying the functional equation of the polygamma function \cite{nbs} we 
obtain the Proposition.

When applying Eq. (A.1), it is useful to keep in mind a relation \cite{grad,sri}
between the polygamma function and the Hurwitz zeta function: $\psi^{(n)}(x)
=(-1)^{n+1}n!\zeta(n+1,x)$.  In particular, when $x=1/2$ or $x=1$ in the
latter relation, values of the Riemann zeta function appear.  In fact,
another starting point for evaluating integrals similar to (A.1) is to
use integral representations for $\zeta(s)$ \cite{nbs,ivic,sri,titch} in terms 
$P_1(x)+1/2$.




\pagebreak

\end{document}